\documentclass[twocolumn,showpacs,amsmath,floatfix,prl,aps]{revtex4}
\usepackage{graphicx}

\def\v#1{\mbox{\boldmath $#1$}}
\newcommand{\Id}[1] {\int \! \! {\rm d}^3 #1}

\renewcommand{\vr} {{\bf r}}

\newcommand{\vA} {{\bf A}}
\newcommand{\vB} {{\bf B}}

\begin{document}
\title{Comparison of exact-exchange calculations for solids in 
current-spin-density- and spin-density-functional theory}
\author{S. Sharma$^{1,2}$}
\email{sangeeta.sharma@physik.fu-berlin.de}
\author{S. Pittalis$^2$}
\author{S. Kurth$^2$}
\author{S. Shallcross$^3$}
\author{J. K. Dewhurst$^4$}  
\author{E. K. U. Gross$^2$}
\affiliation{1  Fritz Haber Institute of the Max Planck Society, Faradayweg 4-6, 
D-14195 Berlin, Germany.}
\affiliation{2 Institut f\"{u}r Theoretische Physik, Freie Universit\"at Berlin,
Arnimallee 14, D-14195 Berlin, Germany}
\affiliation{3 Department of Physics, Technical University of Denmark,\\
Building 307, DK-2800 Kgs. Lyngby}
\affiliation{4 School of Chemistry, The University of Edinburgh, 
Edinburgh EH9 3JJ.}
\begin{abstract}
The relative merits of current-spin-density- and spin-density-functional 
theory are investigated for solids treated within the exact-exchange-only 
approximation.
Spin-orbit splittings and orbital magnetic moments are determined at zero 
external magnetic field. We find that for magnetic (Fe, Co and Ni) 
and non-magnetic (Si and Ge) solids, the exact-exchange current-spin-density 
functional approach does not significantly improve the accuracy of the 
corresponding spin-density functional results.
\end{abstract}
\maketitle

In the past 30 years, several generalizations of density functional theory 
(DFT) have been proposed. In the early 70's, DFT was extended to spin-DFT 
(SDFT) \cite{BarthHedin:72} by including the spin magnetization as basic quantity 
in addition to the density. This allows for coupling of the spin degrees of 
freedom to external magnetic fields and produces better results for spontaneously 
spin-polarized systems using approximate functionals. Adding yet another density, 
the paramagnetic current, leads to the framework of current-SDFT (CSDFT)
\cite{VignaleRasolt:87,VignaleRasolt:88}. CSDFT includes the coupling of the 
external magnetic field, through its corresponding vector potential, to the 
orbital-degrees of freedom \cite{VignaleRasolt:88}.

SDFT has been enormously successful in predicting the magnetic properties of
materials. This success can be attributed to 
the availability of exchange correlation (xc) functionals which, even 
though originally designed for non-magnetic systems, could be systematically
extended  to the spin polarized case. The most popular of these functionals are
the local spin density approximation (LSDA) and the generalized gradient
approximation (GGA).
CSDFT, on the other hand, has not enjoyed the same attention mainly because
of problems which arise in the
extension of LSDA and GGA to include the paramagnetic
current density \cite{SkudlarskiVignale:93,takada98,higuchi06}. 
Exposing the
homogeneous electron gas to an external magnetic field leads to the appearance 
of Landau levels which, in turn, give rise to derivative discontinuities in 
the resulting xc energy density. Using this quantity
to construct (semi-)local functionals then automatically 
leads to local discontinuities in the corresponding xc
potentials, which are then awkward to use in practical calculations. 

Such problems can be avoided with the use of orbital functionals and this fact,
coupled with the success of these functionals for SDFT
calculations, has led to recent interest in orbital functionals
for CSDFT \cite{RohraGoerling:06,PittalisKurthHelbigGross:06,HelbigKurthPittalisEsaGross,
RohraEngelGoerling,Heaton-BurgessAyersYang:07}. 
The results from these works have shown mixed success.
A modified version of the original CSDFT 
\cite{Bencheikh:03} 
lead to promising results for
spin-orbit induced splittings of bands in solids, such as Si and Ge
\cite{RohraEngelGoerling}.
In contrast it was found that for open-shell atoms and quantum dots the 
difference between SDFT and CSDFT results was minimal
\cite{PittalisKurthHelbigGross:06,HelbigKurthPittalisEsaGross}. Similarly,
calculations for solids using a local vorticity 
functional \cite{EbertBattoclettiGross:97} and for quantum dots using a 
LSDA-type xc functional \cite{Esa:03} could not establish the superiority of 
CSDFT over SDFT.

In this work we present a systematic comparison of the relative merits of 
CSDFT and SDFT for solids.
Since the Kohn-Sham (KS) system in CSDFT reproduces the current of the
interacting system, one would expect differences between SDFT and CSDFT 
results for orbital magnetic moments (which can be 
directly derived from the current). With this in mind we calculate the 
orbital magnetic moment of the spontaneous magnets Fe, Co and Ni. Since CSDFT 
is believed to improve the spin-orbit induced band splitting in 
the non-magnetic semiconductors Si and Ge \cite{RohraEngelGoerling} 
it makes these materials interesting candidates for a study of the 
differences between the two approaches.

Following Vignale and Rasolt \cite{VignaleRasolt:87,VignaleRasolt:88}, the
ground state energy of a (non-relativistic) system of interacting electrons
in the presence of an external magnetic field ${\bf B}_0({\bf r}) = \nabla
\times {\bf A}_0({\bf r})$ can be written as functional of three independent densities
the particle density $\rho({\bf r})$, the magnetization density ${\bf m}({\bf r})$
and the paramagnetic current density ${\bf j}_p({\bf r})$. This functional
is given by
\begin{widetext}
\begin{eqnarray}\label{eq:Etot}
E[\rho,{\bf m},{\bf j}_p] &=& T_s[\rho,{\bf m},{\bf j}_p] + U[\rho]
+E_{\rm xc}[\rho,{\bf m},{\bf j}_p] + \int \rho({\bf r}) v_{\rm 0}({\bf r})
\,d^3r \\ \nonumber
&-& \int {\bf m}({\bf r})\cdot{\bf B}_{\rm 0}({\bf r})\,d^3r
+ \frac{1}{c} \int {\bf j}_p({\bf r})\cdot{\bf A}_{\rm 0}({\bf r})\,d^3r
+ \frac{1}{2 c^2} \int \rho({\bf r}) {\bf A}_{\rm 0}^2({\bf r})\,d^3r,
\end{eqnarray}
where $ T_s[\rho,{\bf m},{\bf j}_p]$ is the kinetic energy functional of
non-interacting electrons, $U[\rho]$ is the Hartree energy, and
$E_{\rm xc}[\rho,{\bf m},{\bf j}_p]$ is the exchange-correlation energy.
Minimization of Eq.~(\ref{eq:Etot}) with respect to the three basic densities
leads to
the Kohn-Sham (KS) equation which reads
\begin{equation}\label{KS}
\left[ \frac{1}{2}
\left( -i \nabla + \frac{1}{c} {\bf A}_{\rm s}({\bf r})
\right)^2 +v_{\rm s}({\bf r}) - \mu_B \boldsymbol\sigma 
\cdot{\bf B}_{\rm s}({\bf r})
\right] \Phi_j({\bf r})=\varepsilon_j \Phi_j({\bf r}) \; .
\end{equation}
Here $\boldsymbol\sigma$ is the vector of Pauli matrices and 
the $\Phi_i$ are spinor valued wave functions. 
The effective potentials $v_{\rm s}$, ${\bf B}_{\rm s}$ 
and ${\bf A}_{\rm s}$ are such that the ground-state densities 
$\rho$, ${\bf m}$ and ${\bf j}_p$ of the 
interacting system are reproduced. These effective potentials are given by 
\begin{equation}
v_{\rm s}({\bf r})=v_{\rm 0}({\bf r}) + v_{\rm H}({\bf r}) 
+  v_{\rm xc}({\bf r})+\frac{1}{2 c^2} \left( {\bf A}_{\rm 0}^2({\bf r}) 
- {\bf A}_{\rm s}^2({\bf r}) \right), \, \,
{\bf B}_{\rm s}({\bf r})={\bf B}_{\rm 0}({\bf r}) 
+ {\bf B}_{\rm xc}({\bf r}), \, \, 
{\bf A}_{\rm s}({\bf r})={\bf A}_{\rm 0}({\bf r})+{\bf A}_{\rm xc}({\bf r}).
\end{equation}
Here, $v_{\rm 0}$ is the external electrostatic potential and 
$v_{\rm H}({\bf r})=\int\rho({\bf r}')/|{\bf r}-{\bf r}'|\,d^3r'$ is the 
Hartree potential. 
The xc potentials are given as functional derivatives of the 
xc energy with respect to the corresponding conjugate 
densities which can be obtained from KS wave functions using
the following relations
\begin{equation}
\rho({\bf r}) = \sum_{i=1}^{\rm occ} \Phi_i^{\dagger}({\bf r}) \Phi_i({\bf r}), \,\,
{\bf m}({\bf r}) = - \mu_B \sum_{i=1}^{\rm occ} \Phi_i^{\dagger}({\bf r}) 
\boldsymbol\sigma \Phi_i({\bf r}), \, \,
{\bf j}_p({\bf r}) = \frac{1}{2 i} \sum_{i=1}^{\rm occ} \left\{ 
\Phi_i^{\dagger}({\bf r}) \nabla \Phi_i({\bf r}) -
\left[ \nabla \Phi_i^{\dagger}({\bf r})\right] \Phi_i({\bf r}) \right\} 
\end{equation}
where the sum runs over the occupied orbitals.
For practical calculations, an approximation for the xc
energy functional $E_{\rm xc}[\rho,{\bf m},{\bf j}_p]$ has to be adopted. 
Here we concentrate on approximations of the xc functional 
which explicitly depend on the KS orbitals and therefore only 
implicitly on the densities. 
Such orbital functionals are usually treated within the framework of the 
so-called Optimized Effective Potential (OEP) method 
\cite{SharpHorton:53,TalmanShadwick:76,GraboKreibichKurthGross:00,engel93} where 
the xc potential is obtained as solution of the OEP integral equation. 
Recently, the OEP method has been generalized to non-collinear
SDFT \cite{SharmaDewhurstDraxlKurthPittalisGrossShallcrossNordstroem} and 
CSDFT \cite{PittalisKurthHelbigGross:06,HelbigKurthPittalisEsaGross}. 
Another generalization of the OEP method in the context of a spin-current 
DFT (SCDFT) based on a different choice of densities has also been put 
forward \cite{RohraGoerling:06}. In the present work the formalism of
Refs. (\onlinecite{PittalisKurthHelbigGross:06}) and (\onlinecite{HelbigKurthPittalisEsaGross}) is 
used and the corresponding OEP equations can be put in a compact form as 
\begin{equation}\label{oep}
\sum_{k=1}^{\rm occ} \Phi_k^{\dagger}(\vr)\Psi_k(\vr) + h.c. = 0, \,\,
-\mu_B\sum_{k}^{\rm occ} \Phi_k^{\dagger}(\vr)\v\sigma\Psi_k(\vr) + h.c.= 0,\,\,
\frac{1}{2i}\sum_{k}^{\rm occ}
\left\{\Phi_k^{\dagger}(\vr)\nabla\Psi_k(\vr)
-\left[\nabla\Phi_k^{\dagger}(\vr)\right]\Psi_k(\vr)\right\} + h.c. = 0 \;,
\end{equation}
where the so-called orbital shifts 
\cite{GraboKreibichKurthGross:00,KuemmelPerdew:03} are defined as 
$\Psi_k(\vr)=\sum_{j}^{\rm unocc}\frac{\Phi_j(\vr)\Lambda_{kj}}{\varepsilon_k-\varepsilon_j}$,
here the summation runs over the unoccupied states and
\begin{equation}
\Lambda_{kj}= \Id{r'} \biggl(v_{\rm xc}(\vr')\rho_{kj}({\bf r'})
+\frac{1}{c}\vA_{\rm xc}(\vr')\cdot{\bf j_p}_{kj}({\bf r'})
-\vB_{\rm xc}(\vr')\cdot{\bf m}_{kj}({\bf r'}) -\Phi_j^{\dagger}(\vr')
\frac{\delta E_{\rm xc}}{\delta\Phi_k^{\dagger}(\vr')} \biggl)
\end{equation}
where
$\rho_{kj}({\bf r})=\Phi_j^\dag({\bf r})\Phi_k({\bf r})$,
${\bf m}_{kj}({\bf r})=-\mu_B \Phi_j^\dag({\bf r})\boldsymbol\sigma\Phi_k({\bf r})$ and
${\bf j_p}_{kj}({\bf r})=\frac{1}{2i} \left\{ \Phi_j^\dag({\bf r})\nabla\Phi_k({\bf r}
- \left[ \nabla\Phi_j^\dag({\bf r}) \right] \Phi_k({\bf r}) \right\}.$
\end{widetext}

Eq.~(\ref{oep}) has a structure very similar to the OEP 
equations for non-collinear SDFT differing only by the redefinition of the 
matrix $\Lambda$, which now also contains an extra term depending upon the current
density and its conjugate field. Due to their similar structure the CSDFT OEP 
equations are solved by generalizing the `residue algorithm', successfully applied
to solve the non-collinear SDFT 
equations \cite{SharmaDewhurstDraxlKurthPittalisGrossShallcrossNordstroem,KuemmelPerdew:03,KuemmelPerdew:03-2}.
The only difference in the case of CSDFT is introduction of an additional residue
coming from the third OEP equation in Eq. (\ref{oep}).
In the present work we have used the exchange-only exact-exchange (EXX) 
functional to solve the OEP equations. 
The EXX (gauge invariant) energy functional is the Fock exchange energy but evaluated with 
KS spinors 
\begin{equation}\label{eq:EXX}
E^{\rm EXX}_{\rm x}[\{\Phi_i\}]\equiv -\frac{1}{2}\int\int\sum_{i,j}^{\rm occ}
\frac{\Phi_i^\dag({\bf r})
\Phi_j({\bf r})\Phi_j^\dag({\bf r}')\Phi_i({\bf r}')}{|{\bf r}-{\bf r}'|}
\,d^3r\,d^3r' \; .
\end{equation}

In order to keep the numerical analysis as accurate as possible, in the present 
work all calculations are performed using the state-of-the-art full-potential 
linearized augmented plane wave (FPLAPW)
method \cite{Singh}, implemented within the EXCITING code \cite{exciting}.
The single-electron problem is solved using an augmented plane wave basis
without using any shape approximation for the effective potential.
Likewise, the magnetization and current densities and their conjugate fields
are all treated as unconstrained vector fields throughout space.
The deep lying core states (3 Ha below the Fermi level) are
treated as Dirac spinors and valence states as Pauli spinors.
To obtain the Pauli spinor states, the Hamiltonian containing
only the scalar fields is diagonalized in the LAPW basis: this is the
first-variational step. The scalar states thus obtained are then used as
a basis to set up a second-variational Hamiltonian with spinor degrees
of freedom, which consists of the first-variational eigenvalues along
the diagonal, and the matrix elements obtained from the external and
effective vector fields in Eq. (\ref{KS}). This is more efficient than simply
using spinor LAPW functions, but care must be taken to ensure there are
a sufficient number of first-variational eigenstates for convergence of
the second-variational problem. Spin-orbit coupling is also included at this
stage.

As was shown above for CSDFT, the magnetic field couples not
only to spin but also to the orbital degrees of freedom through the
vector potential. This makes CSDFT specifically important 
for magnetic materials and particularly interesting for their orbital
properties.
By analogy with SDFT, one might expect that the introduction of the 
paramagnetic current density gives an improvement in properties
such as orbital moments and spin-orbit induced band splitting, which are related 
to this new basic variable. However, within the framework of existing 
functionals it is yet to be established conclusively that CSDFT performs better
than SDFT for these properties.
The recent development of the OEP method both for SDFT and CSDFT allows for a direct 
comparison of these two approaches for the same xc functional, namely EXX.

\begin{table}[t]
\begin{tabular}{c|c|ccc|c}
\hline\hline 
      &      &       & SDFT  &       & CSDFT  \\ 
Solid & Exp. & LSDA  & GGA   & EXX   & EXX    \\ \hline  
Fe    & 0.08 & 0.053 & 0.051 & 0.034 & 0.034  \\ 
Co    & 0.14 & 0.069 & 0.073 & 0.013 & 0.013  \\ 
Ni    & 0.05 & 0.038 & 0.037 & 0.029 & 0.029  \\ \hline
      &      & 36.2  & 36.7  & 63.4  & 63.4   \\ \hline\hline
\end{tabular}
\caption{\label{tab1} Orbital magnetic moments for bulk Fe, Co and Ni in
$\mu_{B}$. The experimental data are taken from Ref. (\onlinecite{Stearns:87}). 
The final row lists the average percentage deviation of the numerical results 
from the experimental value.}
\end{table}

The orbital moments of spontaneous magnets Fe, Co and Ni, in the absence of 
external magnetic fields and with spin-orbit coupling included, 
are presented in Table~\ref{tab1}. For SDFT, the LSDA, 
GGA and EXX functionals are used, while for CSDFT the values are obtained 
using the EXX functional. It is clear from Table~\ref{tab1} that there is \emph{no
difference} between the results obtained using EXX-CSDFT and EXX-SDFT.
Formally, the ${\bf j}_p$ determined from SDFT
does not correspond to the true paramagnetic current density of the fully
interacting system.
Nevertheless, it is standard practice to compute the orbital magnetic moment 
{\bf L}, like those listed in Table~\ref{tab1}, which is related to ${\bf j}_p$
from the 
KS orbitals by the relation
${\bf L}=\frac{1}{2}\int {\bf r}\times {\bf j_p}({\bf r}) d^3r$.
The fact the EXX-SDFT and EXX-CSDFT orbital moments are so close may be
viewed as a post-hoc justification of this practice for magnetic metals.
It should 
also be noted that in comparison to experiments the EXX results 
are significantly worse than their LSDA and GGA counterparts.
One reason, of course, is the fact that LSDA and GGA also 
include correlation in an approximate way which is neglected completely 
within the EXX framework.

In a recent work \cite{RohraEngelGoerling} it is shown that the use of 
the EXX functional in the framework of SCDFT,
improves the spin-orbit induced splitting of 
the bands in semiconductors. Unfortunately, it is not 
clear if this improvement is due to the use of different 
functionals (going from LSDA to EXX), or due to the use of an extra density 
when going from SDFT to CSDFT.
This has motivated us to compare CSDFT and SDFT results for this quantity 
using the same functional in both cases.
We have determined the value of this splitting for solid Si and Ge and the 
results are presented in Table~\ref{tab2}. While the EXX functional
significantly improves the agreement with experimental values, there
is almost no change on going from SDFT to CSDFT. Thus the improvement
is solely due to the orbital based functional. We also note 
that the EXX-CSDFT results of Ref. (\onlinecite{RohraEngelGoerling})
are significantly different from ours, and in much worse agreement
with experiments. This might be due to the use of pseudopotentials in
the previous work. In this respect it is worth noting that EXX derived KS
energy gaps also show significant differences depending on whether
an all-electron full-potential or pseudopotential method is used
\cite{SharmaDewhurstDraxl:05}.

\begin{table}[t]
\begin{tabular}{c|c|ccc|cc}
\hline\hline 
 Symmetry           &      &       & SDFT  &       & CSDFT             \\ 
 point              & Exp. & LSDA  & GGA   & EXX   & EXX$^{\rm p}$ & EXX$^{\rm o}$ \\ \hline 
Ge $\Gamma_{7v-8v}$ & 297  & 311   & 296   & 291.3   & 289  & 258.1    \\ 
Ge $\Gamma_{6c-8c}$ & 200  & 229.7 & 220   & 201.3   & 199  & 173.3    \\ 
Si $\Gamma_{25v}$   & 44   & 50    &  58   &  42.5   & 45.5 &  42.5    \\ 
\hline
                    &      & 9.5   & 14.0  & 2.0     & 2.2  & 10.5     \\
\hline\hline
\end{tabular}
\caption{\label{tab2} Spin-orbit induced splittings for bulk Ge and Si in meV. 
The experimental data is taken from 
Ref. (\onlinecite{O.Madelung:04}). EXX$^{\rm p}$ are results of the present work 
and EXX$^{\rm o}$ are results from Ref. (\onlinecite{RohraEngelGoerling}). The final 
row lists the average percentage deviation of numerical results from the 
experimental value.}
\end{table}

The paramagnetic current density of Ge for LSDA, GGA, EXX-SDFT
and EXX-CSDFT is plotted in Fig. 1.
Ge is chosen as an example since the 
spin-orbit induced splitting is largest for this system and, unlike in the case 
of metallic orbital moments, this quantity does show some difference on going from SDFT 
to CSDFT. We immediately notice that there is no significant qualitative
difference between the LSDA and GGA currents.
There are, however, pronounced
differences in the current density between LSDA/GGA and EXX-(C)SDFT: the current
in the latter case being smaller and more homogeneous than that of the former.
This is an interesting finding since it indicates the tendency of (semi-) local
functionals towards higher values of the paramagnetic current density.
It is worthwhile noting previous EXX-(C)SDFT results for open-shell atoms in which 
it was found \cite{PittalisKurthHelbigGross:06} that this effect was even more
pronounced and lead to vanishing currents.

Even though the EXX-SDFT current is considerably lower in
magnitude than that of EXX-CSDFT and also has a less symmetric structure, the
spin-orbit splittings for the two cases are almost the same.
Similar conclusions regarding the total energies were
also drawn for quantum dots in external magnetic fields studied using 
EXX \cite{HelbigKurthPittalisEsaGross} and other functionals  of the current 
density \cite{Esa:03}.
From Fig. 1 it is also clear that one of the major effects of using the OEP method 
and of using  ${\bf j}_p$ as an extra density is to change the local structure of the
paramagnetic current, which in turn suggests that quantities 
depending on local properties of the currents, such as chemical shifts, might exhibit 
larger differences in the two approaches. Such calculations \cite{lee95} of 
chemical shifts, performed using \emph{local} functionals, found that for molecules 
this is not the case. The effect of the EXX functional on these shifts may be an 
interesting subject for future investigations.

\begin{figure}[ht]
\centerline{\includegraphics[width=\columnwidth,angle=0]{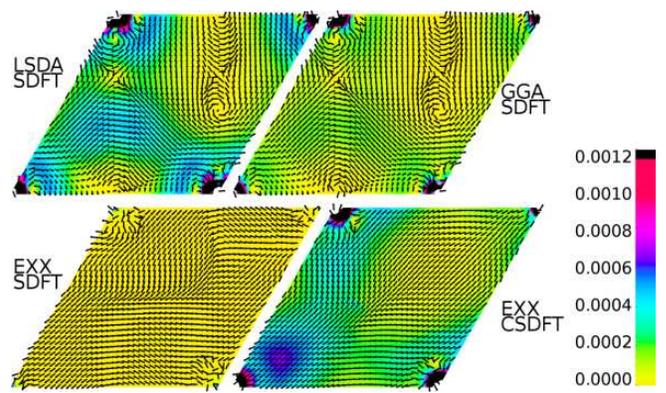}}
\caption{(Color online)Paramagnetic current density for Ge, in the [110] plane, 
calculated using the SDFT and CSDFT. Arrows indicate the direction and information 
about the magnitude (in atomic units) is given in the colour bar.}
\end{figure}

To summarize, in this work we have presented EXX- SDFT and CSDFT 
calculations for solids.
The orbital magnetic moments of Fe, Co and Ni and the spin-orbit induced band splitting of 
Si and Ge are computed. Our analysis shows only minor differences between EXX- CSDFT 
and SDFT results. The spin-orbit induced band splittings in EXX 
calculations are in rather good agreement with experiments, while the results for 
the orbital moments are worse than the LSDA or GGA values.
This highlights the importance of proper treatment of 
correlations for the accurate determination of the orbital moments. 

We acknowledge Deutsche Forschungsgemeinschaft (SPP-1145) and NoE NANOQUANTA 
Network (NMP4-CT-2004-50019) for financial support.


\end{document}